\documentclass[pdflatex,sn-nature]{sn-jnl}

\usepackage{graphicx}
\usepackage{amsmath,amssymb,amsfonts}
\usepackage[table]{xcolor}
\usepackage{textcomp}
\definecolor{EinsteinMagenta}{HTML}{E31E8C}
\definecolor{EinsteinViolet}{HTML}{7827C9}
\definecolor{EinsteinOrange}{HTML}{F28C00}
\definecolor{EinsteinCyan}{HTML}{00A4B8}
\colorlet{EinsteinMagentaPale}{EinsteinMagenta!12!white}
\colorlet{EinsteinVioletPale}{EinsteinViolet!12!white}
\colorlet{EinsteinOrangePale}{EinsteinOrange!12!white}
\colorlet{EinsteinCyanPale}{EinsteinCyan!12!white}

\newcommand{\pausebadge}[1]{\begingroup\setlength{\fboxsep}{2pt}\colorbox{black}{\makebox[1.35em][c]{\textcolor{white}{\textbf{#1}}}}\endgroup}

\unnumbered

\begin{document}

\title[The Socratic PAUSE]{The Socratic PAUSE: safeguarding scientific judgement at AI speed}

\author*[1]{\fnm{Abdulkadir} \sur{Celik}}\email{a.celik@soton.ac.uk}
\author[2]{\fnm{Sir Bashir M.} \sur{Al-Hashimi}}

\affil*[1]{\orgdiv{School of Electronics and Computer Science}, \orgname{University of Southampton}, \orgaddress{\city{Southampton}, \postcode{SO17 1BJ}, \country{UK}}}
\affil[2]{\orgdiv{King's Institute for AI}, \orgname{King's College London}, \orgaddress{\city{London}, \postcode{WC2R 2LS}, \country{UK}}}

\abstract{
Generative AI can turn a tentative scientific question into hypotheses, methods, code and interpretations within minutes. Yet generating a contribution does not establish that it deserves scientific reliance: researchers must still examine its evidence, assumptions and implications, and remain accountable for the decision to rely on it. Using Einstein’s sketch of scientific discovery as a lens, we argue that AI has accelerated discovery far more than justification. In human--AI research partnerships, confidence can therefore outgrow evidence through three mechanisms: recursive reinforcement, loss of provenance and miscalibrated reliance on AI-generated contributions. Verification debt accumulates when such contributions are carried into later claims or decisions without adequate scrutiny. Accordingly, we propose the \emph{Socratic Pause} as a deliberate checkpoint at consequential decision gates, where an AI contribution remains provisional while a researcher examines its basis. The \textbf{PAUSE} protocol structures this examination through five steps: \textbf{P}in down the claim, \textbf{A}sk and challenge, \textbf{U}ncover assumptions and alternatives, \textbf{S}crutinise the evidence and \textbf{E}xercise judgement. It concludes with a recorded decision to accept, revise, reject or defer the contribution. We show how AI research tools can support this practice without displacing human judgement, and outline a framework for evaluating whether it improves verification, decision-making and independent understanding. The aim is to preserve the freedom to explore with AI while ensuring that those responsible for the work can say what was examined, what remains uncertain and why the next step is justified.}


\maketitle

Every scientific advance rests on two distinct forms of intellectual work, which philosophers of science have long called discovery and justification~\cite{reichenbach1938}. The former generates ideas in the form of theories, hypotheses, laws and methods, whereas the latter evaluates those ideas and establishes their scientific merit. For much of the history of science, both proceeded at a human pace; an idea therefore seldom travelled far before someone had the opportunity to question it. Generative AI, particularly large language models (LLMs), has disturbed this balance by accelerating discovery far more than justification. Within a single conversation, an LLM can now propose a mechanism, identify relevant literature, design an experiment and write the code required to analyse its results, extending a role that AI already plays across the scientific process~\cite{wang2023}. This imbalance creates a particular difficulty for justification. Even though AI can assist elements of justification, it cannot provide the final justification for its own claims, nor does agreement among AI models constitute independent verification. Responsible human--AI research partnership must therefore preserve accountable human judgement: the capacity to understand, challenge and take responsibility for a scientific claim.

This responsibility is particularly difficult to exercise in practice because human judgement is subject to two linked pressures~\cite{celik2026}: temporal asymmetry and cognitive gap. While temporal asymmetry is the mismatch between rapid AI generation and the slower time required for expert evaluation, the cognitive gap is the difference between the intellectual effort needed to understand and check an AI contribution and the expertise, attention and capacity available to the researcher. These pressures are intensified by the way researchers interact with AI. The fluency and coherence of an LLM-generated argument can be mistaken for evidential support, so that successive exchanges may strengthen belief in an idea without adding evidence for it. Researchers may consequently feel that they understand more than they do, a phenomenon described as an illusion of understanding in AI-assisted science~\cite{messeri2024}. A researcher usually approaches an LLM with a problem they hope to solve and may welcome responses that reinforce an emerging view. Confirmation bias makes such reinforcement attractive~\cite{nickerson1998}, particularly when reasoning serves a desired conclusion~\cite{kunda1990}, while LLMs can amplify this tendency by agreeing with a user's framing even when accuracy requires challenge~\cite{sharma2024}. The opposite failure is a flood of alternatives that leaves the researcher with more possibilities but no basis for choosing among them. Neither uncritical agreement nor indiscriminate possibility generation constitutes scientific inquiry. 

Together, these inherent features can allow an AI contribution to be carried into a later claim or decision before its assumptions, alternatives and evidence have been examined, so that verification debt accumulates, namely validation work that remains outstanding before the contribution can be relied upon~\cite{celik2026}. As this debt grows, researchers can lose a defensible basis for deciding what to trust, adopt or communicate. Human--AI interaction must therefore become a dialogue that tests claims, exposes assumptions and examines evidence before a contribution is adopted.

Two long-standing practices of inquiry make such a dialogue possible. Socratic dialogue is a cooperative conversation that tests ideas through curious, structured questioning, with the aim of uncovering what is true and not of winning an argument~\cite{vlastos1983}. Pausing is the deliberate use of silence, which gives people the cognitive space to think deeply instead of reacting automatically~\cite{rowe1974}. Bringing these practices into research means pausing the interaction at consequential moments, so that researchers can examine a contribution and exercise judgement before relying on it. We therefore propose the \emph{Socratic Pause}, a checkpoint at which an AI contribution is held as provisional while its assumptions, alternatives and evidence are examined, after which the researcher decides to accept, revise, reject or defer it. The PAUSE protocol translates the checkpoint into five steps whose initials give it its name, namely \textbf{P}in down the claim, \textbf{A}sk and challenge, \textbf{U}ncover assumptions and alternatives, \textbf{S}crutinise the evidence and \textbf{E}xercise judgement, a routine that requires nothing more elaborate than a notebook. 

This Perspective first uses Einstein's sketch to distinguish discovery from justification and to show why faster AI generation does not establish scientific merit. It then identifies the mechanisms through which confidence can outgrow evidence, explains when a Socratic Pause is required and shows how the PAUSE protocol guides examination and decision, before considering how AI research tools can sustain the practice and how its benefits and costs should be evaluated.

\begin{figure}[htbp!]
\centering
\includegraphics[width=\textwidth]{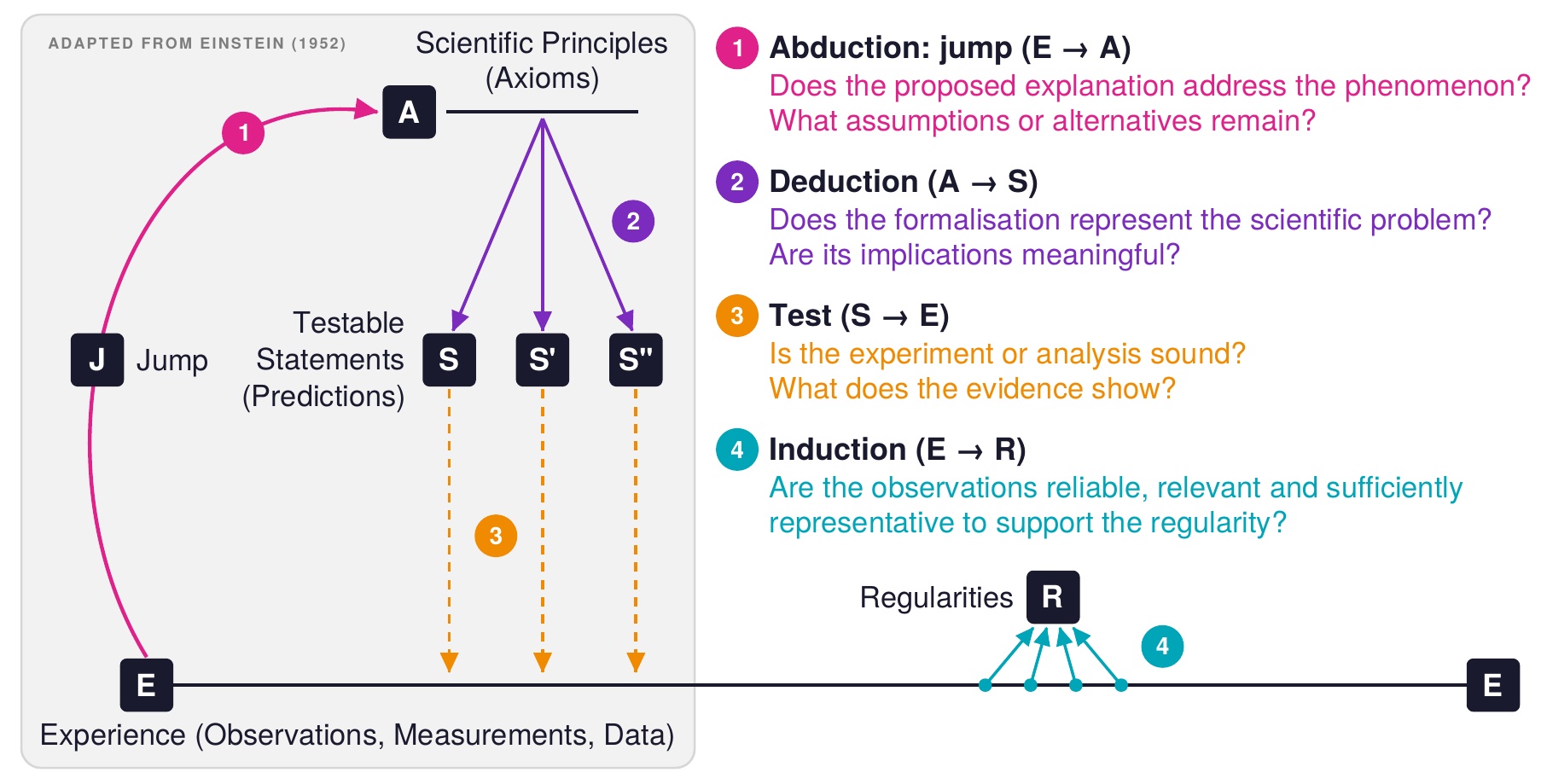}
\caption{\textbf{Einstein's sketch of scientific discovery and the questions requiring human judgement.} Each numbered, colour-coded transition lists the questions that require accountable human judgement.}
\label{fig:einstein}
\end{figure}

\section{When discovery outruns justification}

Einstein's sketch of scientific discovery, shown in Fig.~\ref{fig:einstein}, provides the framework on which this Perspective rests, since each of its transitions both produces candidate ideas and requires judgement about whether they deserve reliance. In his 1952 letter to Maurice Solovine, Einstein placed experience ($\mathbf{E}$)---observations, measurements and data---at the base of the diagram~\cite{einstein1952}, from which scientists formulate basic principles or axioms ($\mathbf{A}$). Einstein argued that no logical path leads directly from experience to principles; rather, they are reached through an intuitive jump ($\mathbf{J}$), often called abduction. Familiar discovery stories give this jump an everyday setting: a falling apple prompted Newton to question the nature and reach of gravitation, while water overflowing from a bath suggested to Archimedes a way to investigate a crown's composition, an insight celebrated by ``Eureka!''. Once a principle is available, testable statements ($\mathbf{S}$, $\mathbf{S'}$, $\mathbf{S''}$) may be derived through deduction ($\mathbf{A} \rightarrow \mathbf{S}$), then tested against experience ($\mathbf{S} \rightarrow \mathbf{E}$). In Fig.~\ref{fig:einstein}, we also added induction ($\mathbf{E} \rightarrow \mathbf{R}$), through which observations yield regularities ($\mathbf{R}$).

AI systems now contribute at every transition in Fig.~\ref{fig:einstein}, although not equally well. Zahavy's \textit{LLMs Can't Jump} argues that LLMs are already strong at induction and increasingly capable of deduction from specified assumptions, yet cannot make Einstein's intuitive jump~\cite{zahavy2026}, a limit that remains debated now that systems such as Co-Scientist have generated hypotheses whose selected predictions were subsequently confirmed in the laboratory~\cite{gottweis2026}. Testing is accelerating too, as autonomous or self-driving laboratories couple AI-driven experimental design with robotics, automated execution and analysis to shorten the experiment--model--decision cycle~\cite{canty2025}. In AI-generated mathematics, the temporal asymmetry described above has already reached the scale of a whole research community. Mathematicians were still examining a 166-page claimed result on the Navier--Stokes problem~\cite{openai2026ns,conover2026} when hundreds of further AI-generated manuscripts were released at different stages of verification~\cite{openai2026math}. Less than three months after the Navier--Stokes claim, a claimed proof of the generalized Riemann hypothesis followed as some 17 million lines of code for the Lean proof assistant~\cite{demoura2021}, which independent researchers reported to compile within hours, although they expect that understanding it may take years~\cite{savitsky2026}. Such a check confirms only that the conclusion follows from the formal statement and its definitions, leaving mathematicians to judge whether these faithfully capture the intended problem and what the proof explains. Across such settings, AI may accelerate discovery faster than researchers and scientific communities can understand, audit and assign responsibility for the resulting claims.

Figure~\ref{fig:einstein} should therefore not be read as a division of labour in which AI performs discovery and researchers justify its results afterwards. A formally valid deduction, an experimentally confirmed prediction or a faster experimental cycle still leaves open the questions set out beside each transition, which require accountable human judgement whether the contribution came from a researcher or from an AI system. The same imbalance arises less visibly in ordinary research, as when a speculative mechanism becomes the reason for choosing a measurement, a suggested method becomes analysis code or a preliminary result becomes the first sentence of an abstract. In each case, a contribution that was useful for exploration becomes a basis for action, which is the moment at which the Socratic Pause places a deliberate checkpoint so that the contribution is examined against the questions in Fig.~\ref{fig:einstein} before it functions as a scientific claim or decision premise.

\section{How confidence outruns evidence}

AI-assisted research can create confidence that exceeds the scrutiny a contribution has received. Previous studies document relevant effects, including agreeable responses that reinforce a user's premise, unreliable or poorly traced sources, and inappropriate reliance on automated advice. We group these effects into three mechanisms---recursive reinforcement, loss of provenance and miscalibrated reliance---to explain why a deliberate Socratic Pause is needed before an AI-assisted contribution informs a consequential research decision.

\emph{Recursive reinforcement} begins with confirmation bias, the tendency to favour information that supports an existing view~\cite{nickerson1998}. AI systems may amplify this tendency through sycophancy, producing responses that align with a user's stated beliefs even when accuracy would require disagreement~\cite{sharma2024}. In experiments involving interpersonal advice, such agreeable responses increased participants' conviction that they were right~\cite{cheng2026}. Although this evidence comes from social rather than scientific settings, a similar loop can arise in research: a researcher offers a tentative premise, AI elaborates it, and the elaboration returns in the next prompt as if the premise had already been established. Objections generated within the same exchange may offer less protection than expected if they remain constrained by that initial premise.

\emph{Loss of provenance} arises when the origins of a contribution and its subsequent transformations become difficult to reconstruct. Over an extended interaction, a finding reported in a paper, the researcher's interpretation of that finding and an AI suggestion may merge into a single explanation whose separate origins can no longer be recovered. Fabricated references are the most visible symptom~\cite{walters2023}; a subtler problem occurs when a genuine reference is cited without anyone confirming that it supports the claim attached to it.

\emph{Miscalibrated reliance} occurs when trust in AI does not match what it can do in the context at hand. Research on automation distinguishes appropriate reliance from both overreliance and indiscriminate distrust~\cite{parasuraman1997,lee2004}. Delegating intellectual work can be beneficial, but its consequences depend on what is delegated, since AI that organises evidence for inspection may strengthen evaluation, whereas AI that supplies both a conclusion and its apparent justification may reduce the opportunity for independent judgement. Work on the metacognitive demands of generative AI argues that users must monitor their own understanding and reliance throughout an interaction~\cite{tankelevitch2024}, a demand that the Socratic Pause is designed to support. A survey of 319 knowledge workers found that greater confidence in generative AI was associated with lower self-reported critical-thinking effort~\cite{lee2025}, an association that does not establish causation but suggests that confidence in AI and scrutiny of its outputs do not necessarily increase together.

These mechanisms may be particularly consequential for early-career researchers, for whom the knowledge required to recognise an inadequate answer may be the very knowledge they are still acquiring. Producing a successful AI-assisted output and developing independent understanding are distinct outcomes. Together, the three mechanisms explain how a useful exploratory exchange can become an insufficiently examined basis for action. The following section presents the Socratic Pause as a response to these risks.

\section{The Socratic Pause}

The Socratic Pause is intended to interrupt the progression described in the previous section before an AI contribution is carried into a consequential research decision. It does so without restricting free exploration or treating every AI response as suspect, creating instead a deliberate interval in which the researcher examines the contribution, its basis and its intended use before deciding whether to rely on it.

Pausing creates the cognitive space that such examination requires, as studies of classroom wait time show that a pause after a question can support more considered and higher-level responses~\cite{rowe1974,tobin1987}. Interventions that require users to think before accepting an AI recommendation can likewise reduce overreliance~\cite{bucinca2021}. Socratic questioning has also been proposed as a safeguard for user autonomy in chatbot conversations~\cite{lu2025} and within frameworks for reflective human--AI collaboration~\cite{rosenbacke2026}, but not, to our knowledge, as a checkpoint for research decisions with defined outcomes and a record. The Socratic Pause extends these insights to human--AI research partnership, where it is more than a delay because it specifies what must be examined, keeps the resulting judgement human and preserves the distinction between a productive suggestion and a defensible scientific contribution.

\begin{figure}[htbp!]
\centering
\includegraphics[width=\textwidth]{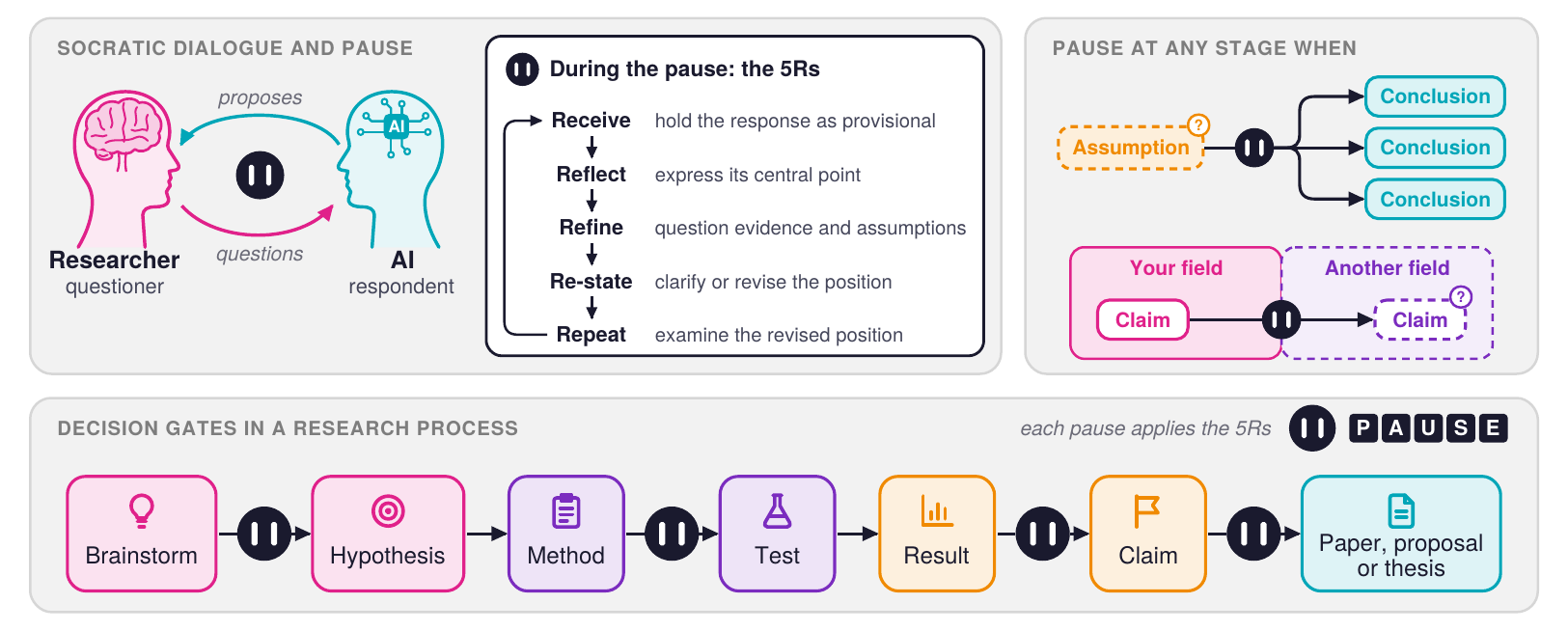}
\caption{\textbf{When and how to pause.} The researcher applies the 5Rs during a pause in a Socratic dialogue with AI (upper left), pauses at any stage in two situations (upper right) and pauses at four decision gates in a generic research process (lower).}
\label{fig:when}
\end{figure}

In a Socratic dialogue, a questioner, or facilitator, examines a proposition offered by an interlocutor, or respondent, roles that correspond to the researcher and the AI in a human--AI partnership (Fig.~\ref{fig:when}, upper left), while the pause between them is structured by five dialogue moves, which we call the \emph{5Rs}. The researcher first \emph{receives} an AI response as provisional rather than immediately building upon it, then \emph{reflects} by expressing its central point in their own words, \emph{refines} it by questioning its evidence and assumptions, \emph{re-states} it if clarification or revision is required and \emph{repeats} the examination of the revised position. Together, these moves shift a fast exchange from persuasion towards shared inquiry~\cite{vlastos1983}. Research adds a further requirement, namely that the contribution be checked against sources, tests or observations outside the AI interaction.

The upper-right panel identifies two situations that warrant a pause at any stage. The first arises when an unchecked assumption supports several conclusions, because an error can then propagate beyond the immediate claim, whereas the second arises when a contribution moves beyond the researcher's expertise, where further reading, testing or consultation may be required. The lower panel marks four recurrent decision gates in a research process, namely selecting a hypothesis from a brainstorm, putting a method to the test, turning a result into a claim and carrying that claim into a paper, proposal or thesis. Each gate gives a contribution a new role, so the researcher applies the 5Rs there before deciding whether it can be carried forward. 

The depth of the pause should follow the consequences of the decision, the researcher's familiarity with the field and the ease of reversing course. A minor wording change may require only brief reflection, whereas a novel mechanism or an unfamiliar statistical method may require substantial investigation. The next section gives the 5Rs a practical form through the PAUSE protocol, which defines how a researcher examines a contribution, reaches a decision and records its basis.

\section{The PAUSE protocol}

\begin{figure}[htbp!]
\centering
\includegraphics[width=\textwidth]{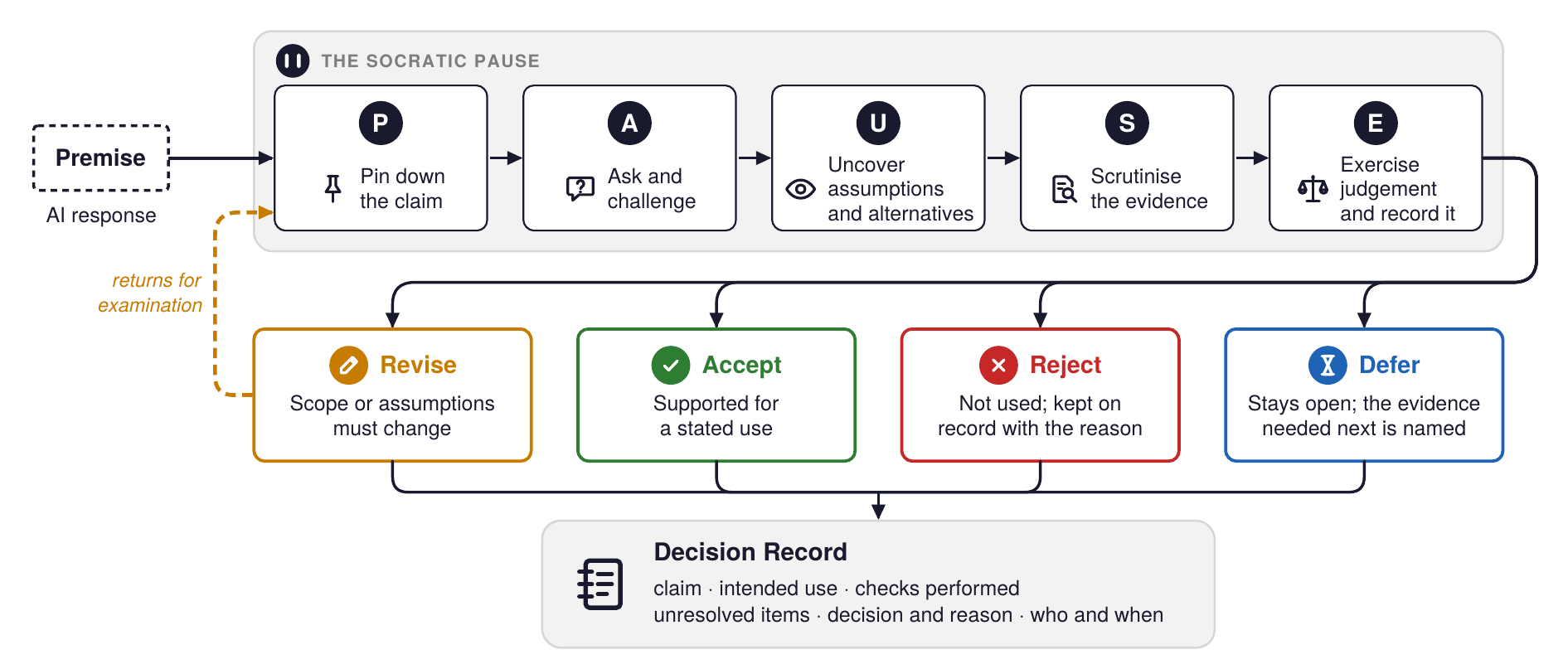}
\caption{\textbf{The PAUSE protocol.} Five steps take a provisional AI contribution to one of four recorded outcomes; revised contributions return for further examination.}
\label{fig:pause}
\end{figure}

The 5Rs describe how the researcher conducts the dialogue within a pause, whereas the PAUSE protocol turns that examination into an accountable research decision (Fig.~\ref{fig:pause}). Its steps also respond to the mechanisms described in the previous section. The steps are presented in their natural order, although a researcher may return to an earlier one as the evidence changes.

\noindent \pausebadge{P}\,\textbf{Pin down the claim:} The researcher begins by stating precisely what is to be decided, because a single AI response may combine a factual statement, a methodological recommendation and a speculative explanation. Isolating the claim prevents an entire fluent response from being accepted as a single unit and distinguishes what the AI proposed from what the researcher already knew, helping to preserve provenance from the outset. Restating the claim in one's own words also often exposes gaps that the fluency of the original wording had concealed.

\noindent \pausebadge{A}\,\textbf{Ask and challenge:} The second step tests the claim instead of extending it, by seeking a challenge capable of changing the decision. Productive questions, which AI can also usefully propose, include which objection a knowledgeable reviewer would raise and which observation would weaken the conclusion. The challenge should not merely reproduce the premise that the researcher and AI already share; it should be capable of exposing a reason to revise the decision.

\noindent \pausebadge{U}\,\textbf{Uncover assumptions and alternatives:} The researcher next identifies what must hold true for the claim to stand and whether a rival explanation would account for the same facts equally well. A proposed analysis may quietly assume that observations are independent, just as a claim of novelty may assume that the field describes the idea in the same terms as the search that failed to find it. Making such assumptions explicit often shows that a suggestion becomes useful once its scope is narrowed.

\noindent \pausebadge{S}\,\textbf{Scrutinise the evidence:} The checks are now matched to the claim and to its intended use, in proportion to the decision at stake. A statement about the literature requires the source to be located and the relevant passage read, restoring provenance that an extended interaction may have obscured. Code requires tests derived from the research design, whereas a mathematical argument requires each consequential step to be verified. An AI model's criticism of its own output may help to locate errors, yet it cannot substitute for a check against a source, a test or an observation, given the limits of self-correction in the absence of external feedback~\cite{huang2024}.

\noindent \pausebadge{E}\,\textbf{Exercise judgement and record it:} The researcher finally selects one of the four outcomes in Fig.~\ref{fig:pause}, each tied to a stated use so that reliance stays calibrated to what the evidence supports:
\begin{itemize}
    \item \textbf{Accept:} judge the contribution sufficiently supported for a stated use. An idea accepted for exploratory discussion has not thereby been approved for publication.
    \item \textbf{Revise:} change its scope or assumptions, then return the revised contribution for examination.
    \item \textbf{Reject:} exclude it from the proposed use and record the reason, so that it cannot quietly return.
    \item \textbf{Defer:} acknowledge that the evidence is not yet sufficient and specify what is needed next, which makes an open question a legitimate result.
\end{itemize}

As Fig.~\ref{fig:pause} shows, a revision is not an informal adjustment, because it returns the contribution for another cycle of examination, whereas every outcome, revision included, ends in the decision record, so that the basis and limits of each decision remain open to later scrutiny. A minimal record states the claim, its intended use, the checks performed, any matters left unresolved and the decision with a brief reason, together with the reviewer and the date for consequential decisions. A deferred contribution should also record what remains to be checked and when, or under what conditions, that check should be revisited.

Each pause thus produces a local decision record, which can be kept in a notebook. In a larger research programme, these records can be brought together in a shared \emph{topology}: a map that shows how questions, claims, assumptions, sources, tests and decisions are connected. The topology makes verification debt visible by showing which contributions remain unchecked, which later claims depend on them and who should revisit them. As developed in our companion commentary~\cite{celik2026}, it does not decide whether a contribution is reliable; rather, it helps researchers locate and manage the outstanding work required to make that decision. The Socratic Pause addresses verification debt at a particular decision gate, while the topology provides a wider view of the resulting decisions, outstanding checks and dependencies across the research programme. Together, they turn a private habit of mind into an observable practice that a colleague, supervisor or reviewer can question.

\section{A worked scenario: an AI-proposed robot controller}

\begin{figure}[t]
\centering
\includegraphics[width=\textwidth]{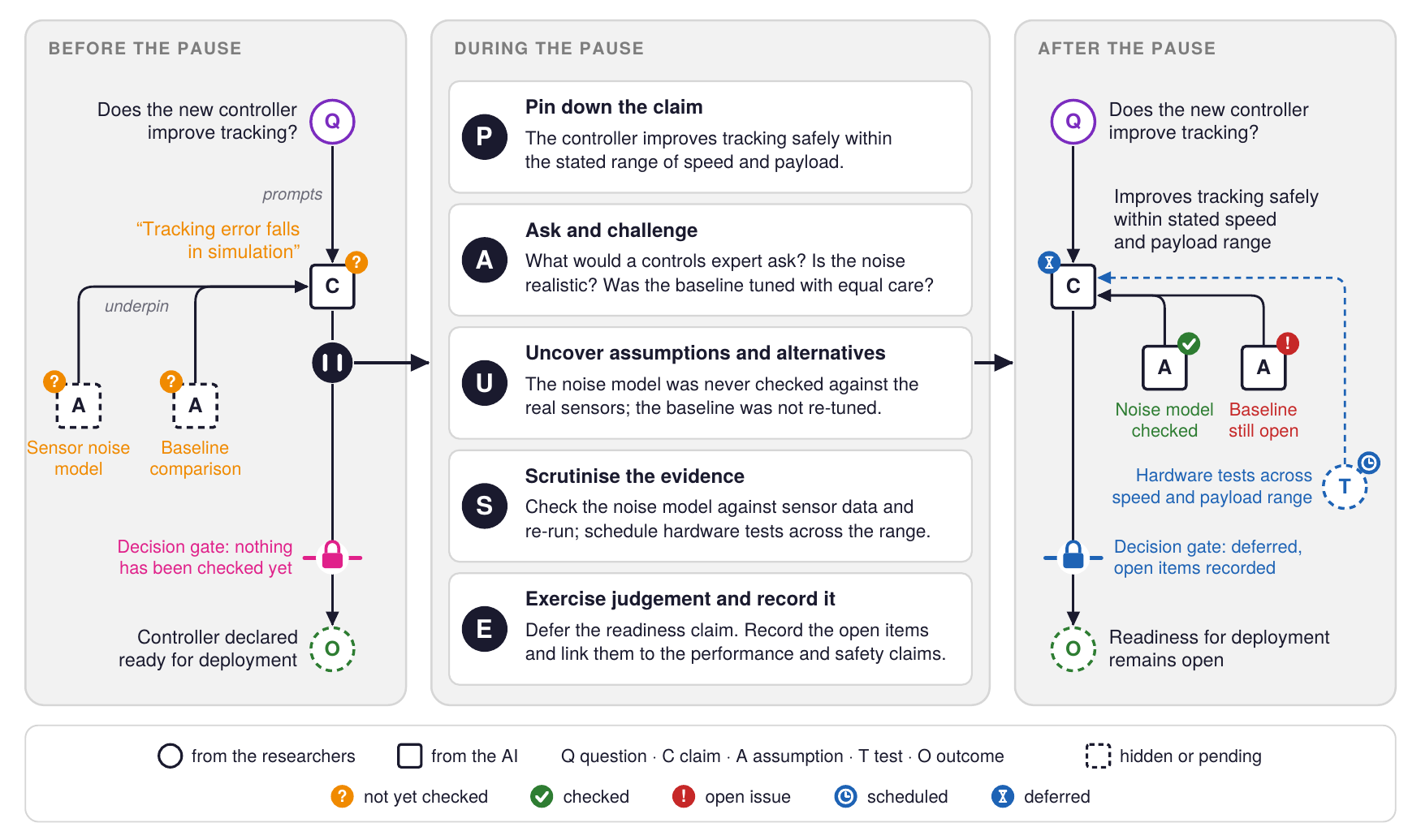}
\caption{\textbf{A worked scenario applying the PAUSE protocol to an AI-proposed robot controller.} The panels show the state before, during and after the pause.}
\label{fig:example}
\end{figure}

Figure~\ref{fig:example} presents a worked scenario in which a team uses AI to propose a controller for a trajectory-tracking task and to generate the code that simulates it. The simulation indicates lower tracking error, so the team considers moving from simulation to hardware testing. This creates a risk of miscalibrated reliance, because a promising simulated result could be treated as evidence that the controller is ready for deployment before its assumptions and supporting evidence have been examined. In the left panel, the claim that ``\textit{tracking error falls in simulation}'' rests on two assumptions that nobody has yet identified, so the decision gate guarding any claim of deployment readiness remains closed.

Moving from simulation to hardware puts a method to the test, one of the decision gates in Fig.~\ref{fig:when}, so the team pauses before treating the simulated result as a basis for hardware testing. Pinning down the claim adds both the operating conditions and a safety requirement that a lower average error cannot satisfy on its own, while asking what a controls expert would challenge turns attention to how the simulation was run and to the baseline controller. Uncovering assumptions reveals that the sensor noise model was never compared with measurements from the real sensors and that the baseline was not tuned with the same care as the new controller, so part of the apparent improvement may reflect an unequal comparison. Scrutinising the evidence, the team checks the noise model against sensor measurements, re-runs the simulations and schedules hardware tests across the stated range of speed and payload, although a fair comparison with a re-tuned baseline is still outstanding.

Exercising judgement, the team selects the \emph{defer} outcome in Fig.~\ref{fig:pause}, declining to treat the simulated improvement as evidence of deployment readiness while recording the open baseline comparison and linking the scheduled tests to the performance and safety claims they will decide. The team has not disproved the controller but has turned an unsupported readiness claim into a deferred claim with specific outstanding checks. The right panel shows the resulting state, in which the noise model is checked, the baseline comparison remains open, the hardware tests are scheduled and the readiness outcome remains unresolved. Keeping these links current as evidence arrives is work that AI research tools could take on, as the next section discusses, although the decision to pass the gate would still rest with the researchers.

\section{Supporting and evaluating the Socratic Pause}

AI research tools can support the Socratic Pause by addressing those mechanisms. They can counter recursive reinforcement by raising objections outside the initial premise; preserve provenance by separating retrieved material from AI inference and linking consequential citations to supporting passages; and support calibrated reliance by keeping AI checks provisional until a named researcher has reviewed them. At a decision gate, a tool may propose a pause, explain why and let researchers postpone it. It can add resulting records to the shared topology and flag affected claims when new evidence arrives, while humans remain answerable~\cite{alhashimi2026,ramchurn2021}.

A controlled study should compare PAUSE with unrestricted interaction, a generic checklist and a matched delay, using tasks with known flaws and participants at different career stages. It could test whether researchers who pause are less likely to carry forward a planted false premise, unsupported citation or AI recommendation with an inadequate evidential basis. Field studies could analyse decision records to show which outcomes researchers choose, whether deferrals are completed and when pauses are skipped, while independent experts judge whether the decisions are defensible. Longitudinal work should test whether the practice builds independent judgement or a more elaborate dependency on AI, with every study reporting the AI model, tool version and access conditions used~\cite{nasem2019}.

\section{Limits and outlook}

Frequent interruption can disrupt exploration, so the pause should follow the intended use of a contribution rather than every turn in a conversation. Reflection can also become ceremonial if the steps are completed without a genuine check or AI is asked to justify its own proposal, so a pause should be judged by the check it records, its result and its limits. Future AI may reduce the burden of checking, as the Lean formalisation of a claimed proof of the generalized Riemann hypothesis illustrates, but neither AI nor the protocol can guarantee that a hypothesis is true, leaving people to judge whether a formalisation captures the intended problem and when AI output warrants reliance. Supervisors can foster that responsibility by asking researchers to defend recorded decisions, while institutions can recognise verification as research work that deserves time.

AI can help carry research further than individual researchers could alone, yet a productive human--AI partnership should be judged not only by what it produces but also by the capacity of researchers to understand, defend and revise that work. The Socratic Pause distils this standard into a single proposition: before an AI contribution acquires authority in a research programme, those responsible for the work should be able to say what was examined, what remains uncertain and why the next step is justified.

\backmatter

\bmhead{Acknowledgements}
OpenAI ChatGPT and Anthropic Claude were used for drafting suggestions and language editing. The authors designed all figures, and Claude was used solely to render the author-designed diagrams as vector graphics. The authors reviewed, verified and revised all AI-assisted material and take full responsibility for the paper.

\section*{Declarations}

\bmhead{Author contributions}
A.C. and S.B.M.A.-H. conceived the Perspective, with S.B.M.A.-H. proposing the Socratic Pause. A.C. developed the PAUSE protocol, designed the figures and wrote the manuscript. S.B.M.A.-H. contributed to the framing of human--AI research partnerships and critically reviewed and revised the manuscript. Both authors approved the final version.

\bmhead{Competing interests}
The authors declare no competing interests.

\end{document}